%% file: main.tex
\documentclass[sigconf,lettersize,9pt,dvipsnames]{acmart}

\usepackage[utf8]{inputenc}
\usepackage[ruled,vlined]{algorithm2e}
\usepackage{amsmath}
\usepackage{enumitem}
\usepackage{cleveref}
\usepackage{amsthm}
\usepackage{listings}
\usepackage{mathtools}
\usepackage{xcolor}
\usepackage{pdfpages}
\usepackage[normalem]{ulem}
\usepackage{tikz}
\usepackage{multirow}
\usetikzlibrary{automata, positioning, arrows}
\tikzset{
    ->,  
    >=stealth', 
    node distance=1.5cm, 
    every state/.style={thick, fill=gray!10,minimum size=0pt}, 
    initial text=$$, 
}
\usepackage{syntax}
\usepackage{stmaryrd}
\usepackage{graphicx}
\usepackage{bbold}
\usepackage{xcolor}
\usepackage{color}
\usepackage[font={footnotesize}]{caption}
\usepackage[font={footnotesize}]{subcaption}

\usepackage{mdframed}
\newmdenv[
  topline=false,
  bottomline=false,
  skipabove=\topsep,
  skipbelow=\topsep
]{siderules}

\definecolor{eclipseStrings}{RGB}{42,0.0,255}
\definecolor{eclipseKeywords}{RGB}{127,0,85}
\colorlet{numb}{magenta!60!black}

\newcommand{\lstbg}[3][0pt]{{\fboxsep#1\colorbox{#2}{\strut #3}}}
\lstdefinelanguage{diff}{
  basicstyle=\ttfamily\small,
  morecomment=[f][\lstbg{red!20}]-,
  morecomment=[f][\lstbg{green!20}]+,
  morecomment=[f][\textit]{@@},
}
\lstdefinelanguage{json}{
    basicstyle=\footnotesize\ttfamily,
    commentstyle=\color{eclipseStrings}, 
    stringstyle=\color{eclipseKeywords}, 
    numbers=left,
    numberstyle=\scriptsize,
    stepnumber=1,
    numbersep=8pt,
    showstringspaces=false,
    breaklines=true,
    frame=lines,
    string=[s]{"}{"},
    comment=[l]{:\ "},
    morecomment=[l]{:"},
    literate=
        *{0}{{{\color{numb}0}}}{1}
         {1}{{{\color{numb}1}}}{1}
         {2}{{{\color{numb}2}}}{1}
         {3}{{{\color{numb}3}}}{1}
         {4}{{{\color{numb}4}}}{1}
         {5}{{{\color{numb}5}}}{1}
         {6}{{{\color{numb}6}}}{1}
         {7}{{{\color{numb}7}}}{1}
         {8}{{{\color{numb}8}}}{1}
         {9}{{{\color{numb}9}}}{1}
}

\usetikzlibrary{shapes,arrows}

\definecolor{lb}{rgb}{.71,.84,.97}
\definecolor{db}{rgb}{.6,.84,.97}		

\definecolor{whitestwhites}{rgb}{.94, .98, .98}
\definecolor{almostwhite}{rgb}{.78, .93, .92}
\definecolor{bb}{rgb}{.43, .83, .79}

\tikzset{
purplebox/.style={
  rectangle,
  inner sep=5pt,
  text width=24mm,
  align=center,
  draw=black,
  fill=almostwhite  
  }
}

\tikzset{
pinkbox/.style={
  rectangle,
  inner sep=5pt,
  text width=22mm,
  align=center,
  draw=black,
  fill=bb,
  }
}

\tikzset{
bluebox/.style={
  rectangle,
  inner sep=5pt,
  text width=24mm,
  align=center,
  draw=black,
  fill=whitestwhites
  }
}

\tikzset{
greenbox/.style={
  rectangle,
  inner sep=2pt,
  text width=5mm,
  align=center,
  draw=black,
  fill=bb
  }
}

\tikzset{
nobox/.style={
  inner sep=2pt,
  text width=10mm,
  align=center,
  execute at begin node=\setlength{\baselineskip}{0em}
  }
}
\hypersetup{draft}

\theoremstyle{remark} 
\newtheorem{definition}{Definition}[section]
\newtheorem{remark}{Remark}[section]

\newcommand\prognosis{\textsc{Prognosis}\xspace}
\newcommand\learner{\textsf{Learner}\xspace}
\newcommand\adapter{\textsf{Adapter}\xspace}
\newcommand\mapper{\textsf{Mapper}\xspace}
\newcommand\implementation{\textsf{Implementation}\xspace}
\newcommand\sul{\textsf{SUL}\xspace}
\newcommand\ot{\textsf{Oracle Table}\xspace}

\ccsdesc[500]{Theory of computation~Logic and verification}


\copyrightyear{2021} 
\acmYear{2021} 
\setcopyright{acmcopyright}\acmConference[SIGCOMM '21]{ACM SIGCOMM 2021 Conference}{August 23--27, 2021}{Virtual Event, USA}
\acmBooktitle{ACM SIGCOMM 2021 Conference (SIGCOMM '21), August 23--27, 2021, Virtual Event, USA}
\acmPrice{15.00}
\acmDOI{10.1145/3452296.3472938}
\acmISBN{978-1-4503-8383-7/21/08}

\begin{document}
\title[\prognosis: Closed-Box Analysis of Network Protocol Implementations]{\prognosis: Closed-Box Analysis \\of Network Protocol Implementations}

 \author{Tiago Ferreira}
 \affiliation{%
   \institution{University College London}
 }
 
    \author{Harrison Brewton}
        \affiliation{%
   \institution{University of Wisconsin--Madison}
 }
 
    \author{Loris D'Antoni}
     \affiliation{%
   \institution{University of Wisconsin--Madison}
 }

 \author{Alexandra Silva}
 \affiliation{%
   \institution{University College London}
 }
 
\renewcommand{\shortauthors}{Ferreira et al.}

\begin{abstract}

We present \prognosis, a framework offering automated closed-box learning and analysis of models of network protocol implementations. \prognosis\ can learn models
that vary in abstraction level from simple deterministic automata to models containing data
operations, such as register updates, and can 
be used to unlock a variety of analysis techniques---model checking temporal properties, computing differences between models of two implementations of the same protocol, or improving
testing via model-based test generation. 
\prognosis\ is modular and easily adaptable to different protocols (e.g., TCP and QUIC) and their implementations. 
We use \prognosis\ to learn models of (parts of) three QUIC implementations---Quiche (Cloudflare), Google QUIC, and Facebook mvfst---and use these models to analyze the differences between the various implementations. Our analysis provides insights into different design choices and uncovers potential bugs. Concretely, we have found critical bugs in  multiple QUIC implementations, which have been acknowledged by the developers.
\end{abstract}

\maketitle

\section{Introduction}
\input{tex/introduction}


\section{Architecture}
\label{sec:Architecture}

\input{tex/framework}

\section{Learning Module}
\label{sec:learning}

\input{tex/process}

\input{tex/checking}

\newcommand{\Prob}{\textbf{P}}
\section{Results}
\label{sec:results}
\input{tex/evaluation}

\section{Related Work}\label{sec:related}
Model learning has been applied to a range of communication and security protocols~\cite{aarts-inference-2010,aarts-formal-2013}, including network protocols~\cite{DBLP:conf/nordsec/Ruiter16,tls-paper,tcp-paper,10.1145/2976749.2978383}. \prognosis improves on previous work in several directions: first, the architecture of \prognosis is parametric with respect to reference implementation, 
an aspect that greatly decreases the amount of work and expertise needed to analyze one protocol implementation; second, 
\prognosis allows a user to swap different implementations of  the same protocol seamlessly by requiring only a socket change; third, \prognosis uses synthesis to enrich the models with data and enables a more fine-grained control in the type of properties and details one can analyze. \prognosis is the first model-learning tool used to analyze several QUIC implementations.

Other approaches to analyze QUIC include recent work by \citet{mcmillan}  who manually build a formal specification of the wire protocol in the Ivy language, and then use it for test generation and find a range of bugs in different implementations. In contrast, the approach used in \prognosis automates the building of a finite state model guided by an abstract alphabet. 
The models learned by \prognosis can also be used  for test generation, and crucially one does not need to manually encode the protocol logic, including complex cryptographic  components, to use \prognosis. \citet{mcmillan} specify the full protocol state including the security handshake, with unbounded data. The models we learned with \prognosis abstract away details not covered by the alphabet or the registers we chose to synthesize. Closely related to \cite{mcmillan}, but only for TCP, is the work of \cite{DBLP:journals/jacm/BishopFMNRSSW19} using  symbolic model checking.

Other approaches to analyze protocol implementations include building a {\em correct-by-construction} reference implementation~\cite{everest,delignat-lavaud-security-nodate} and then use it when testing a new implementation. This approach has the advantage of having every component of the protocol in the reference implementation formally verified. However, the amount of manual work to build such implementation is out of reach of most development teams and, moreover, it will be making specific choices in terms of the RFC specification. 

\prognosis is closed-box---it does not assume access to the code of the implementation being analyzed. Open-box approaches to analyze QUIC also appeared in the literature~\cite{10.1145/3284850.3284853,goel-testing-2020}. These works focus on testing protocol compliance using symbolic execution. 
However, subtle bugs related to ambiguities in the RFC or differences between implementations (like the ones detected by \prognosis and described in \Cref{sec:results}) would not be detected easily.

Finally, \prognosis complements differential testing~\cite{mckeeman-differential-1998}: the learned models as well as the \adapter can be used to create high-quality test cases that trigger complex behaviors of the protocols, something that is typically hard in a closed-box setting. Furthermore, \prognosis produces models that can be inspected by protocol designers, something that differential testing cannot help with.

In summary, our work focuses on developing a modular, reusable, and flexible framework, where different formal analysis techniques can easily be made accessible for different protocols and implementations thereof. 

\section{Conclusion}\label{sec:discussion}
\prognosis is a modular framework to automatically learn models of network protocol implementations and analyze them. \prognosis has been successfully used to analyze TCP and QUIC implementations and has found several bugs in different mainstream QUIC implementations. 
One of the key contributions of \prognosis\ is the use of a reference implementation to remove the burden 
of having to implement a protocol logic from the user. This step was required in previous approaches based on model learning. 
Though this contribution means that less experienced users have easier access to our framework, 
\prognosis still requires some knowledge on how to instrument the reference implementation. 

One  direction for further work is investigating whether there are semi-automated ways of aiding the user of \prognosis in finding the key places where the code needs instrumentation. The use of {\em active} model learning is core to \prognosis as the precision of the learned model can be adjusted on demand. However, in cases where access to logs is possible, and to avoid resorting to so many expensive queries, the learning process could be speeded up using a combination of {\em passive} and {\em active} learning. 

The models \prognosis can currently synthesize do not encompass any form of {\em environment quantities}, which are interesting in a networking context to capture e.g., congestion, latency, or memory usage properties. Extending \prognosis to richer quantitative models is perhaps the most challenging yet impactful direction for future work, as it will require significant advances on the design of learning algorithms. Recent developments in active learning of weighted automata~\cite{DBLP:conf/cai/BalleM15,pid} provide a good starting point for this direction. Learning quantitative properties will also require an enriched \adapter that can answer quantitative queries. 

\section{Acknowledgments}
Ferreira and Silva were partially funded by ERC grant AutoProbe (101002697). D'Antoni was partially funded by Facebook research awards, a Microsoft Faculty Fellowship, and an Amazon Research Award. The authors would like to thank Hongqiang Liu for his help in preparing the final version of the paper.

\bibliographystyle{ACM-Reference-Format}
\bibliography{refs.bib}

\begin{center}
	\emph{Appendices are supporting material that has not been peer-reviewed.}	
\end{center}

\input{tex/artifact}
\clearpage
\input{tex/appendix}
\end{document}

%% file: tex/introduction.tex
Implementations of network and security protocols such as TCP/IP, SSH, TLS, DTLS, and QUIC are prominent components
of many internet and server applications. Errors in the implementations of these protocols are common causes of security breaches and network failures (e.g., the Heartbleed OpenSSL vulnerability~\cite{noauthor-nvd-nodate-1} that rendered SSL/TLS connections futile due to leaked secret keys, and others~\cite{noauthor-nvd-nodate-2,noauthor-nvd-nodate-3}). 
 Many approaches have been proposed to verify both specifications and implementations of the above protocols in an attempt to provide safety and correctness guarantees~\cite{tcp-paper,tls-paper,fiterau-brostean-analysis-2020,mcmillan,DBLP:journals/iacr/Delignat-Lavaud20}. 
For example, \citet{everest} are designing a formally verified (using a theorem prover) implementation of the TLS protocol that is provably free of many types of security vulnerabilities.


Most existing approaches suffer from being \textit{monolithic}---small changes
to the implementation require large changes to the verification process---and require \textit{high expertise}---one 
needs to know the protocol very well and have a priori knowledge of what properties the implementation should satisfy.
One way to avoid these limitations 
is to build \textit{models} of 
the implementation, which provide abstractions of the critical components of an implementation and enable a number of powerful analyses such as model-checking and model-based test generation.

For example, instead of directly analyzing the binary or source code of a TCP implementation, one can analyze a model 
that only describes what types of packet flags ($\texttt{SYN}$, $\texttt{ACK}$, etc.) the implementation exchanges during a handshake. 
However, this approach is still monolithic.
First, analyzing the model requires guessing what properties it must satisfy. 
Second, designing models still requires expertise, and whenever the implementation
is modified, one has to manually update the model to retain its faithfulness to the implementation, updating not only the modified sections but also ensuring existing ones are truly unaffected.
This problem is well-known in the model checking literature and recently \citet{mcmillan} showed,
using Microsoft QUIC as an example,
that creating a faithful model of a protocol implementation is a challenging, time-consuming problem that requires
a number of iterations between the model designer and the protocol-implementation developers.

Orthogonally,
several works have shown that  one can often
\textit{learn} a model from implementations  (e.g., for passport~\cite{aarts-inference-2010} and bank card~\cite{aarts-formal-2013} protocols). This idea is called 
{\em model learning}~\cite{vaandrager-model-2017} and it builds on the fact that many classes of finite automata (often used
as models) can be inferred by testing the implementation on a set of traces. 
 \citet{tcp-paper} applied model learning to detect an anomaly in a real TCP implementation, but their system is still monolithic---i.e., it requires one to \textit{manually} design a mapper between the \textit{abstract} traces of the model
and the \textit{concrete} traces of the implementation, a task requiring expert knowledge of the protocol logic. 
Thence, the system of \citet{tcp-paper} is not reusable for different protocols and implementations.

In this paper, we consider the following question:
\begin{center}
\emph{Can we design general and reusable techniques to detect logic errors affecting the interaction with protocol implementations without knowing a priori what properties these errors may violate?}
\end{center}

We present \prognosis, a \textit{modular} and \textit{reusable} framework for learning
and analyzing
 models of network protocol implementations, specialising in bug finding and knowledge acquisition rather than providing verified guarantees. Unlike existing approaches,
 \prognosis can easily be adapted to handle different protocols and protocol implementations, and the programmer \textit{does not} need to manually program the logic to map abstract traces to concrete traces of the protocol implementation, relying instead on a \emph{reference implementation} that they trust by instrumenting its protocol logic components. 
Although any valid implementation of the protocol can be used as a reference implementation, different implementations will vary in their ease of modification depending on how they were designed.
 The \prognosis \adapter, generated from the instrumentation, creates the ideal separation of concerns between implementation-specific details and 
protocol-specific details.
Crucially, the same \adapter can be reused for different implementations of the same protocol with a simple change, as reference implementations are, by design, able to communicate with any other valid implementation of the same protocol. 
Furthermore, \prognosis makes it easy to experiment with different levels of abstraction in the model precision and with several analyses that expose varying types of bugs.

We use \prognosis to analyze 1 implementation of TCP and 4 of QUIC---Quiche (Cloudflare), Google QUIC, Facebook mvfst, and QUIC-Tracker~\cite{piraux-observing-2018}.
For QUIC, we experiment with varying levels of abstractions and learn models that: 
1) only characterize the packet's frames, and
2) concrete data such as packet numbers.
The first type of abstraction allows us to use existing decision procedures to compare whether the models learned from different implementations are equivalent. When dealing with protocols as big as QUIC for example, finding a difference in the models does not necessarily indicate a bug, it can also signal different design decisions as allowed by QUIC's flexible specification. Nonetheless, some of these differences the models capture provide greater insight into the consequences of specific design decisions that are sometimes taken lightly, or sections of the specification that should be stricter. We found one such case which triggered a change to the specification to better formalize the intended behavior.
Abstract models also allow us to verify and test safety and liveness properties. By inspecting counterexamples produced by checking safety properties we identified 3 bugs in QUIC implementations, which have been acknowledged by the developers.

\noindent {\bf Contributions and road map.} In a nutshell, the paper contains the following contributions:

\begin{enumerate}[leftmargin=*]
\item  \prognosis, a framework with a modular architecture that enables reusability: different protocols and protocol implementations can easily be swapped without changes to the learning engine (\Cref{sec:Architecture}). 

\item Configurable levels of abstraction in learned models. This helps with scalability and performance yet does not compromise correctness---once a bug is discovered at the model level, \prognosis creates concrete 
traces to check whether the bug is in the implementation or is a false positive (that can be used to refine the model). (\Cref{sec:learning}). 

\item The various types of model \prognosis can learn and synthesize expose a range of analyses, e.g., model-based testing and decision procedures, which we use in finding bugs. (\Cref{sec:Model-Analysis}). 
 
\item Case studies: a TCP implementation and several QUIC implementations. The TCP case study shows that our framework can recover previous results~\cite{tcp-paper}, albeit with much less manual configuration. The QUIC case study highlights that the framework can be used to uncover bugs as well as to gain insight into different design choices and aid in RFC Improvements (\Cref{sec:results}). The bugs uncovered in two popular implementations of QUIC were confirmed by the developers. One of the bugs (if exploited) could make the QUIC
server vulnerable to DoS attacks and it highlighted that in fact a part of the protocol was totally missing from the implementation.
\end{enumerate} 
We review relevant related work in \Cref{sec:related} and conclude the paper in \Cref{sec:discussion} with a discussion of the limitations of the framework, and directions for further work. 

\smallskip	
\noindent\textit{\textbf{Ethical statement}} Work does not raise any ethical issues.

%% file: tex/framework.tex
In this section, we give an overview of the architecture of our framework. 
\prognosis is comprised of three modules (see Figure~\ref{fig1a}), which are parametrized by inputs provided by the user who is analyzing the protocol. 

\subsubsection*{System Under Learning (\sul)}
\label{sec:sul-short}
The \sul is the protocol implementation we are analyzing---e.g., a TCP server.
This implementation is accessed in a closed-box fashion---i.e., all we assume is that we can send packets to and 
receive packets from it. 
In general,  implementations are complex, and we cannot hope to directly learn models of their entire behaviors.
A user of \prognosis provides an \adapter pair $(\alpha, \gamma)$ containing an abstraction function $\alpha$ that maps concrete packet traces of the \sul 
to simplified abstract traces. For example, the function $\alpha$ might simplify TCP packets to only consider
whether they are of type SYN, ACK, or SYN-ACK.
We also need a concretization function $\gamma$ that maps simplified abstract traces to concrete ones accepted by the \sul. Unlike existing approaches that require the user to explicitly implement this function~\cite{tcp-paper},
\prognosis only requires the user to instrument an existing reference implementation of the same protocol we are analyzing
to produce concrete packet traces from abstract ones. We describe this setup in detail in Section~\ref{sec:sul}.
Now that we have an \adapter that can map between abstract and concrete traces, we have the interface required to 
learn models and perform analysis.

\subsubsection*{Learning Module}
\label{sec:learningmodule}
We use existing active model learning algorithms~\cite{learnlib} to learn a model of the \sul. Active learning algorithms work by directly interacting with the \sul instead of relying on passive data such as logged events or historic data. This allows us to do closed-box learning, where we don't have access to the inner workings of the \sul, and guarantees that the learner is able to get answers to every query it may formulate.

Specifically, we focus on algorithms based on the \emph{Minimally Adequate Teacher Framework}, where the learning algorithm can perform 2 types of queries: membership and equivalence. Both types of queries are further explored in Section~\ref{sec:learning}.

On an abstract level, the models operate over the simplified abstract traces and therefore abstract away much of the complexity of the \sul, allowing the user to focus on the properties of interest.

On a concrete level this involves sending sequences of input packets to the \sul, and registering the response obtained. For example, the learner might ask what happens when a SYN packet is sent and followed by an ACK. The \sul uses its \adapter to, acting as a client, produce and send these packets to the implementation, and return an answer the learner uses to build an accurate model.

In Section~\ref{sec:learning}, we show how \prognosis can learn different models of varying levels of abstraction---from plain deterministic models to models that include registers to store e.g., packet numbers.

We show how \prognosis can be used to learn a detailed model of the TCP 3-way handshake shown in \Cref{TCP-Handshake}. 

\begin{figure}
\centering
	
\scalebox{0.96}{
	    \begin{tikzpicture}
        \node at (0,0) [bluebox] (a) {\large \sc Analysis Module};
        \node at (6,0) [bluebox] (l) {\large \sc Learning Module};
         \node at (3,3) [bluebox] (sul) { \sc System Under Learning (\sul) \phantom{\rule{2cm}{0.48cm}} };
         \node at (-.5,2.5) [nobox] (cq) {\scriptsize Check queries};
         \node at (3,-.275) [nobox] (cq) {\scriptsize Learned Model(s)};
         \node at (.7,.82) [nobox] (cq) {\scriptsize Property Set};
         \node at (5.3,.82) [nobox] (cq) {\scriptsize Abstract Alphabet};
                  \node at (3.65,1.9) [nobox] (cq) {\scriptsize Translation Pair};
                  \node at (6.5,2.5) [nobox] (cq) {\scriptsize Learning queries};
                  \node at (3,.8) [nobox] (cq) {\scriptsize \bf User};
                         \node at (3,2.49) [rectangle,draw=black, fill=almostwhite] (cq) {$(\alpha, \gamma)$};
                      \node at (1.14,0.28) [rectangle,draw=black, fill=almostwhite] (cq) {$\Phi$};
          \node at (4.8,0.27) [rectangle,draw=black,inner sep=2pt, fill=almostwhite] (cq) {$\widehat\Sigma$};
          \node[inner sep=0pt] (user) at (3,1.25)
    {\includegraphics[width=.05\textwidth]{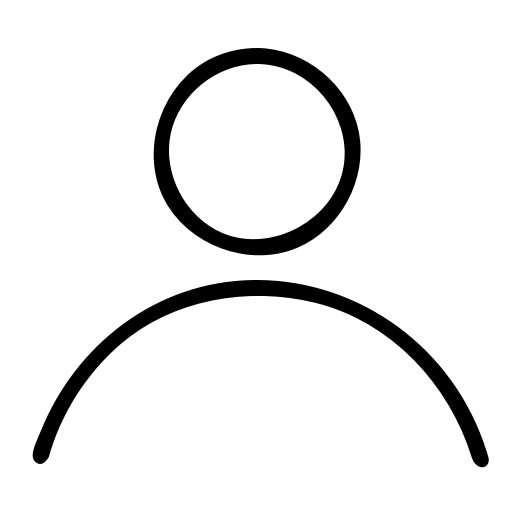}};
                 \draw (l) -- (a);
                  \draw (0,3) -- (sul);
                  \draw (0,3) -- (a);
                  \draw (6,3) -- (sul);
                  \draw (6,3) -- (l);
                   \draw (user) -- (sul);
                   \draw (user) -- (4.8,1.25) -- (4.8,0.5) ;
                   \draw (user) -- (1.2,1.25) -- (1.2,0.5);
    \end{tikzpicture}
    }
\caption{\prognosis' modular architecture.\label{fig1a}}
\end{figure}

\subsubsection*{Analysis Module}

This module enables the use of the learned models to analyze the \sul, using a portfolio of techniques
to unveil complex  bugs and help the user gain insights about the implementation's behavior. 
For example, \prognosis can automatically compare whether the models learned for two different implementations are equivalent (for different notions of equivalence) and also supports simple visualizations of the learned models that allow a user to visually compare two models for differing behaviors. 
In Section~\ref{sec:results}, we show how the automated equivalence check and visualizations aided in detecting anomalies and explaining such anomalies to real developers in our
evaluation. For example, \prognosis could detect that a supposedly variable value being transmitted was actually a constant. 
The full set of analysis exposed by \prognosis is discussed in Section~\ref{sec:Model-Analysis}.

\section{System Under Learning (\sul)}
\label{sec:sul}

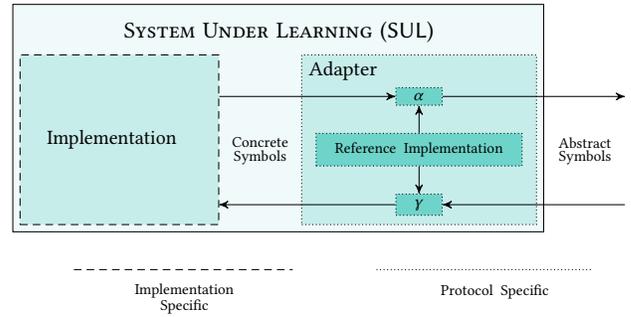
\begin{figure}
\scalebox{0.96}{
    \begin{tikzpicture}
	    \node at (0,0) [bluebox, text width=7cm] (out) {\phantom{\rule{2cm}{2.8cm}}};
	    \node at (0,1.2) [nobox,  text width=7cm] (sul) { \sc System Under Learning (\sul)};
	    
        \node at (-2.2,-0.3) [purplebox, style=densely dashed] (imp) {\phantom{\rule{0cm}{2cm}}};
        \node at (-2.7,-0.3) [nobox] (cg) {\small Implementation};

        \node at (1.95,-0.3) [purplebox, style=densely dotted, text width=29mm] (ada) {\phantom{\rule{0cm}{2cm}}};
        \node at (0.9,0.65) [nobox] (cg) {\small Adapter};        

        \node at (1.95,-0.4375) [pinkbox, style=densely dotted, text width=2.5cm] (ref) {\phantom{\rule{0cm}{0.1cm}}};
        \node at (1.95,-0.4375) [nobox, text width=5cm] (cg) {\scriptsize Reference Implementation};
            
        \node at (1.95,0.30) [greenbox, style=densely dotted] (al) {\scriptsize $\alpha$};
        \node at (1.95,-1.20) [greenbox, style=densely dotted] (ga) {\scriptsize $\gamma$};
        
        \draw (ref.north) -- (al.south);
        \draw (ref.south) -- (ga.north);

		\draw (imp.east|-al.west) -- (al.west);
		\node at (-0.25,-0.45) [nobox] (cq)  {\scriptsize Concrete Symbols};
	    \draw (ga.west) -- (imp.east|-ga.west);

	    \draw (al.east) -- (4.8,0.30);
	    \node at (4.25,-0.45) [nobox] (cq) {\scriptsize Abstract Symbols};
    	\draw (4.8,-1.20) -- (ga.east);
    	         
 		\node at (-1.3,-2.5) [nobox,  text width=3cm] (cq)  {\scriptsize Implementation\ \\ Specific};
 			\draw [style=densely dashed,-] (-2.83,-2.1) -- (0.2,-2.1);
  		
  		\node at (3,-2.4) [nobox, text width=3cm] (cq)  {\scriptsize Protocol Specific};
	  		\draw [style=densely dotted,-] (1.36,-2.1) -- (4.36,-2.1);
 \end{tikzpicture}}
	 \caption{System Under Learning components. \label{fig1b}}
\end{figure}

The \sul has two sub-components: An \implementation we want to analyze and learn a model of, 
and a protocol-specific \adapter that uses a translation pair to transform concrete
packet traces into simplified abstract traces, and vice-versa. The \adapter is the interface with the learning module (Figure~\ref{fig1b}). In fact, the learner is completely oblivious to the existence of concrete traces, and communicates only directly with the \adapter, which then communicates with the \implementation. The learning module (described in  Section~\ref{sec:learningmodule}) will use the abstract traces to build
models of the implementation.

\subsection{\implementation}
The \implementation is the original system we are attempting to learn, for example a running TCP server. In this case we host the implementation being learned, but simply being able to connect to the implementation is enough for \prognosis to learn a model.
At the high level, the implementation takes sequences of inputs in a domain $\Sigma$ 
and produces sequences of outputs in a domain $\Gamma$.

In what follows, we refer to these domains as \textit{alphabets}, the usual terminology in the model learning
literature. We will be using different alphabets that get gradually more abstract: the native alphabet, the concrete alphabet, and the abstract alphabet. Each of these alphabets serves its own purpose, and they are the way in which the \prognosis modules communicate. In general, the input and output alphabets of the implementations are complex packets.
The following example shows the low-level alphabet used in typical network communications.

\begin{figure*}[h]
	\centering
	\includegraphics[width=.2\linewidth]{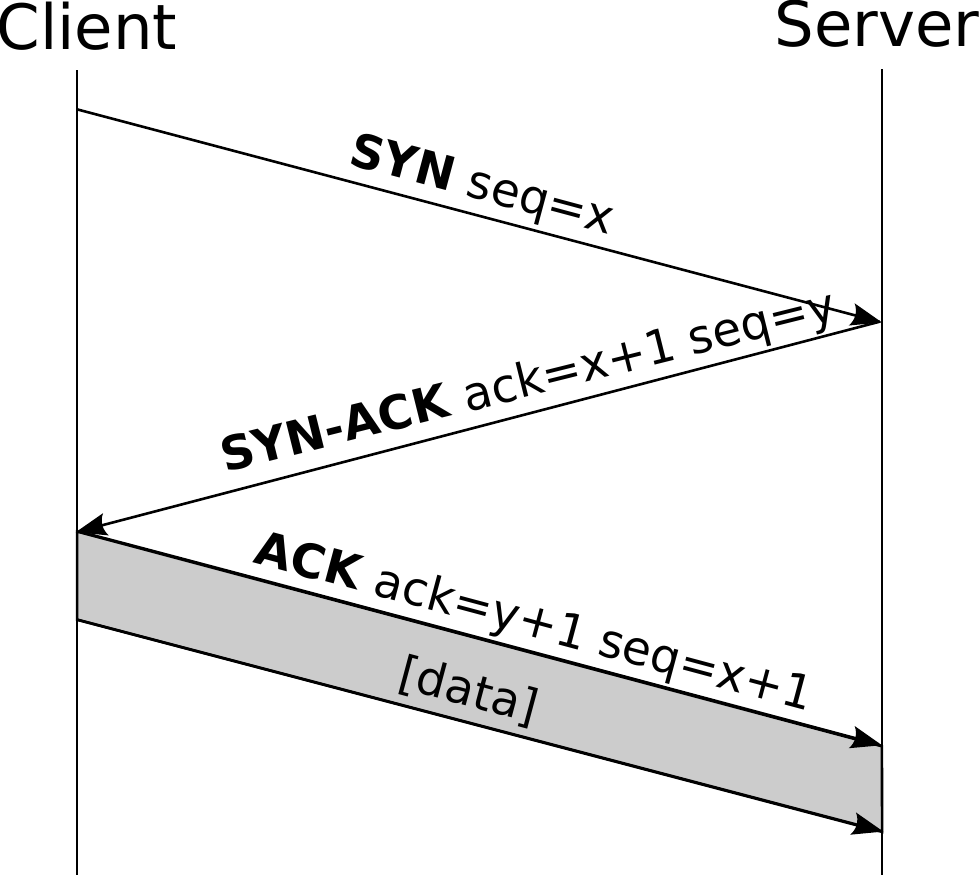} 
	\quad \qquad     
	\begin{tikzpicture}[node distance = 1.2cm, auto, transform shape]
        \node[state, initial] (s0) {$s_0$};
        \node[state, below of=s0] (s1) {$s_1$};
        \node[state, below of=s1] (s2) {$s_2$};
        \draw   (s0) edge[left] node{\texttt{\scriptsize SYN(?,?,0)} / \texttt{\scriptsize ACK+SYN(?,?,0)   }} (s1)
        		(s1) edge[left] node{\texttt{\scriptsize ACK(?,?,0)} / \texttt{\scriptsize NIL}} (s2);
	\end{tikzpicture} \quad \qquad
	\begin{tikzpicture}[node distance = 1.2cm, auto, transform shape]
        \node[state, initial] (s0) {$s_0$};
        \node[state, below of=s0] (s1) {$s_1$};
        \node[state, below of=s1] (s2) {$s_2$};
        \draw   (s0) edge[left] node{\begin{tabular}{c}
				\texttt{\scriptsize SYN(\textnormal{sn},ack,0)} / \texttt{\scriptsize ACK+SYN(seq,\textnormal{r},0)} \\
				\texttt{\scriptsize \textnormal{r} = \texttt{\textnormal{sn}+1}}
				\end{tabular}
				} (s1) 
        		(s1) edge[left] node{\begin{tabular}{c}
				\texttt{\scriptsize ACK(seq,ack,0)} / \texttt{\scriptsize NIL}
				\end{tabular}} (s2);
	\end{tikzpicture}
	\caption{ \quad(a) The TCP 3-way handshake. \quad\quad (b) Learned model (fragment) \quad (c) Synthesized model (fragment) with register {\textnormal{r}} }\label{TCP-Handshake}
\end{figure*}
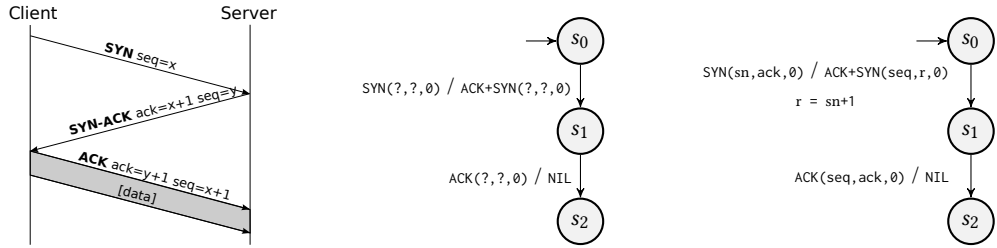

\begin{example}[Native Alphabet]
The {\em native alphabet} of a  TCP implementation consists of all the possible TCP packets in their binary format, with all their complex
fields and restrictions. 
The input and output alphabets are both $\mathbb{2}$, the binary alphabet, and the packets are sequences of these---i.e.,  sequences in $\mathbb{2}^*$. These sequences correspond to the binary representations of TCP packets that will be sent over the wire to the \implementation and received back as a response. 
For example,  in our analysis of  the TCP \implementation in Figure~\ref{TCP-Handshake}, we use the \emph{Scapy} \cite{noauthor-scapy-nodate}  Python library to send arbitrary TCP  packets in $\mathbb{2}^*$ over a socket.
\end{example}

The native alphabet is only useful for creating packets to be communicated, but in general, it is more convenient to use
structured alphabets that capture the fields and values of packets. Even though packets are communicated in binary, these same can be represented as a structured alphabet---e.g., a \textit{JSON} object. 

\begin{example}[Concrete Alphabet]\label{ex:concrete}
In our TCP example, the concrete alphabets can be JSON objects with the following structure:
\begin{lstlisting}[language=json,basicstyle=\ttfamily\scriptsize]
{ "isNull": false,
  "sourcePort": 40965,
  "destinationPort": 44344,
  "seqNumber": 48108,
  "ackNumber": 0,
  "dataOffset": null,
  "reserved": 0,
  "flags": "S",
  "window": 8192,
  "checksum": null,
  "urgentPointer": 0 }
\end{lstlisting}
\end{example}
We will use concrete alphabets---which we below refer to as $\Sigma$ and $\Gamma$---as a machine-readable representation of the native alphabets. 
Concrete alphabets and native alphabets are necessary to trigger concrete executions of the \sul. However, these  alphabets are simply too large (sometimes infinite). When learning a model of an implementation, one has to abstract some parts of the alphabets---e.g.,  specific packet formats and encrypted messages---to make model learning feasible and helpful. For example, while it is not feasible to learn a model describing an exact TCP implementation, it is possible
to learn a model of what types (e.g., \texttt{SYN}, \texttt{ACK}, \texttt{SYN-ACK}) of packets are exchanged during a handshake (Figure \ref{TCP-Handshake}(b)). This abstraction is handled by the \adapter, as described next.

\subsection{\adapter}
\label{sec:adapter}

In the architecture of our system (\Cref{fig1a}), the \adapter, which is part of the \sul (\Cref{fig1b}), is by far the hardest component the user needs to provide. The \implementation itself is provided to us, and the \learner is abstract, but the \adapter has to translate inputs that the \implementation understands to inputs that the \learner can use, and vice-versa. The \adapter has to know the logic of the protocol to produce concrete packets with the right parameters, and it needs to know how to encode concrete symbols into their binary representation to interact with the \implementation, acting as the client in the connection. Instead of requiring the user to provide an implementation of the \adapter from scratch, \prognosis lets the user resort to a \emph{reference implementation} as basis to build the \adapter. Before we describe this key idea, we set some notation.

The \adapter translates concrete input/output traces into simplified traces for which it is feasible to do model
learning. We call the latter abstract traces as they are built from 
\textit{abstract alphabets} $\widehat\Sigma$ and $\widehat\Gamma$, which allow us to focus on crucial aspects by abstracting away details that would make the alphabet otherwise too big or infinite.
Let us now illustrate a candidate abstract alphabet to learn the TCP handshake of \Cref{TCP-Handshake}.

\begin{example}[Abstract Alphabet]\label{ex:abstractalph}
In \Cref{TCP-Handshake}(a) we have packet flags, sequence, and acknowledgment numbers. Because we are only interested in modeling what types of packets are exchanged, our abstract alphabets
$\widehat\Sigma$ and $\widehat\Gamma$
will only contain packet flags.
For example, a packet might have the following structure:
$\texttt{ACK+SYN(?,?,0)}$. Here, each \texttt{?} represents a parameter left unspecified. While it seems unnecessary to have these parameters if they are not used in learning, we will use them in~\Cref{sec:Synthesis} to synthesize a richer model.\end{example}

\begin{remark}[Nondeterminism]\label{remark:nd}
The abstract trace in the previous example does not incorporate the \texttt{seqNumber}.
Because in TCP the Sequence Number is randomly determined at the start of the connection, we could have two different traces that represent the same 3-way TCP handshake. A choice the user has to make when providing an abstract alphabet is what
they plan to model. In this example, we assume the user is trying to learn a model of the 3-way handshake and they know
this process should be deterministic. Therefore, providing elements of the abstract alphabet that not only are irrelevant to
the model but will cause nondeterminism would be a poor choice of abstract alphabet.

\Cref{sec:results} shows how \prognosis provides mechanisms to detect when a choice of abstract alphabet results in nondeterminism.
In \Cref{sec:nondeterminism-bug}, we show a case in which nondeterminism was the result of an undesired 
protocol behavior. 
\end{remark}

Going from a native alphabet, to a concrete alphabet, to an abstract alphabet is somewhat simple as we merely remove information deemed unnecessary for learning. However, these details are essential for the learner to communicate with the \implementation, so this missing information will need to be recovered when sending concrete packets to the \sul.
Learning algorithms (and our analysis) often need to perform queries to the \sul to
decide how the model should be constructed. 
These queries, given an abstract trace $a$, require one to
construct a concrete trace $c$ that is valid in the \implementation and that corresponds to $a$.

Formally, these translations are user-defined functions: An abstraction function $\alpha\colon \Sigma^*\times \Gamma^* \rightarrow \widehat\Sigma^*\times \widehat\Gamma^*$ that maps pairs of input/output traces to abstract traces; and a \textit{concretization function} $\gamma\colon 
\hat\Sigma^*\times \hat\Gamma^*\rightarrow \Sigma^*\times \Gamma^*$ satisfying $\alpha(\gamma(a))=a$. 

As an example of an abstraction function, we could be removing the \texttt{SeqNumber} and \texttt{AckNumber} of TCP Packets: e.g., TCP\{flags: \texttt{SYN}, Seq: \texttt{123}, Ack: \texttt{0}\} would be translated to an abstract TCP Packet TCP\{flags: \texttt{SYN}, Seq: $\top$, Ack: $\top$\}.

Designing the reverse process, that is the concretization function, is a hard task that requires expert knowledge and stands in the way of modularity and reusability. Simplifying this problem is one of the aspects where our work significantly differs from prior work:
 \citet{tcp-paper} have used an architecture reminiscent of ours to learn models of TCP implementations, but require the user to provide a \mapper, which is effectively an implementation of the concretization function $\gamma$. Directly implementing $\gamma$  requires a user to know the protocol logic in detail and to understand what concrete packets are valid and not valid.
Essentially, the user needs to implement part of the protocol implementation itself, which not only is a hard task, but somewhat defeats the point of using a closed-box analysis based on model learning. 
Moreover, this explicit implementation of a concretization function is close to impossible if the protocol relies on a logic of high complexity (as is the case for QUIC), including aspects like key derivation, encryption, or symbols that contain a large number of fields. 

Our solution to implementing concretization functions is inspired by the following common expression in cryptography: \textit{Never roll your own Crypto.}---i.e., because cryptographic algorithms tend to be of great complexity, it is best to use existing implementations that have been widely tested instead of implementing them yourself.
Building on this insight, \prognosis uses the following key idea.

\ \\
\noindent\begin{minipage}{.75\textwidth}
\textbf{\textit{Reference implementation as a concretization oracle.}}\\
\end{minipage}

Instead of manually implementing a concretization function from scratch (i.e., a version of the protocol
logic that can produce concrete traces from abstract ones),
we rely on a given \textit{reference implementation} to provide ground
truth information to the \adapter. Concretely, we rely on the reference implementation to both do the concretization logic and native formatting, as a normal implementation would. 
Given an abstract query symbol $a$ the \adapter needs to find a concrete
packet $c$ to build that matches $a$.
We modify the reference implementation so that it can ``abstractly execute'' abstract packets to identify what
concrete packets they can yield.
While this sounds like a mouthful, it boils down to identifying and using the parts of the reference implementation that operate over the symbols we are interested in. 

The main advantage of using a well-tested reference implementation is that it already encodes the
desired protocol logic---i.e., in what order packets are transmitted, how they are built, etc.
For example, a natural choice for the QUIC reference implementation is QUIC-tracker~~\cite{piraux-observing-2018}, which is widely used by QUIC implementers. 

It is worth noting that this implementation does not have to be 100\% correct. It is merely a reference point for modelling the interaction with the target implementation. While correct reference implementations allow us to narrow detected bugs to the target implementation, finding any modelled bugs is useful as it can detect issues in both implementations.

Naturally, although a reference implementation allows us to communicate with the target implementation, it also has behaviour that is not fit for being modelled. Namely, via our modification of the reference implementation we aim to enforce the following properties:
\begin{description}[style=unboxed,leftmargin=*]
	\item[(1) No unrequested packets are sent to the target Implementation.] We must ensure that any output symbols registered are indeed caused only by the input packets requested. 
	\item[(2) All concrete packets sent match the requested abstract packets.] Concrete packets must fulfil the requested abstract symbols fully. 
	\item[(3) Both reference and target implementations can be reset on request, returning them to their initial state.] The learner must be able to perform a series of independent queries, as such we must have a way to fully reset the connection to its initial state, ready for a new query to be done.
	\item[(4) Concrete packets constructed or received as a response are saved with their abstract counterparts in a historic \ot.] The \ot is a critical data structure used in synthesising more detailed models of the implementation. It is further explored in Section~\ref{sec:Synthesis}.
	\item[(5) Response packets from the target implementation to the reference implementation are abstracted and sent back to the learner.] Just as the adapter must be able to concretise packets, it must be able to abstract them back to the same abstraction level as the original request it received, so that they can be sent back to the learner in a matching abstraction level.
\end{description}

Although the specific ways of achieving these properties vary from an implementation to another, we have found this requirement set is often enough to strike the balance in allowing the learner to interact with the \sul while minimising behaviour introduced by the reference implementation.

In some more complex protocols, it may also be useful to store in a queue formulated payloads that would break rule (1) by reacting to a received response packet, so that they can be sent later if requested by a matching abstract counterpart, as exemplified in Listing~\ref{lst:instrumented1}.

Our modified code will have instrumentations at all the points where packet types and packet frames are 
constructed, and ensure through the instrumented conditional branching that only matching concrete packets are sent to the target implementation. We also add code to be able to hook into functionality allowing us to create packets from scratch.

As an example, consider the \adapter receives an abstract query \texttt{INITIAL[ACK]} for a QUIC packet (where \texttt{INITIAL} is one of the 7 packet types and \texttt{ACK} is one of the 20 frames carried in packets).

Firstly, the queue is searched for a matching packet to be sent. If there is no match for this packet, one is made from scratch, with the valid current state of the reference implementation (Packet Number, connection details, and others), and sent instead. The reference implementation may then receive packets as a response, which are processed by the reference implementation as usual to make sure its state is updated, and if the reference implementation logic would cause a new packet to be sent in response, it is instead stored in our queue waiting for the learner, as demonstrated below.

\begin{lstlisting}[language=diff, frame=lines, label={lst:instrumented1}, caption={Enforcing (1) and (2) in ACK sending logic}, basicstyle=\ttfamily\scriptsize]
  if (packet.shouldBeAcked) {
    // Received a packet to be acked
-   connection.sendPacket(INITIAL, ACK)
+   AckQueue += packet
  }
+ if (requestedInitialAck) {
+   // Abstract Symbol was requested
+   connection.sendPacket(AckQueue.find(INITIAL, ACK))
+ }
\end{lstlisting}

After this, the original response from the target implementation is abstracted, and sent to the learner:

\begin{lstlisting}[language=diff, frame=lines, label={lst:instrumented2}, caption={Implementing (4) and (5) in packet send/receive hooks}, basicstyle=\ttfamily\scriptsize]
+ // Define abstract symbol according to set abstraction.
+ abstractSymbol = newAbstractSymbol(
+   packetType, version, packetNumber, frameTypes)
 (...)	
+ // Save symbol exchange in Oracle Table.
+ oracleTable.addIOs(abstractInputs, abstractOutputs,
+                    concreteInputs, concreteOutputs)
+ // Return abstract response to learner.
+ learner.send(abstractOutputs)
\end{lstlisting}

Finally, at the end of each query the \learner requests that the \sul reset its state for a new query to be made. In the specific case of QUIC, it is enough to reset the \adapter (client), and start a new connection to the \implementation:

\begin{lstlisting}[language=diff, frame=lines, label={lst:instrumented2}, caption={Implementing (3) in main \adapter body}, basicstyle=\ttfamily\scriptsize]
+ // Define reset routine.
+ func reset() {
+   agents.stopAll()
+   connection.close()
+   connection = newConnection(conncetion.getConfig())
+   agents = defaultAgentsWithConnection(connection)
+ }
\end{lstlisting}

Most importantly, by using a reference implementation we can rely on it to maintain the desired state in between multiple packets---i.e., when a new abstract symbol request comes in, we can simply resume from the state we left at the previous abstract symbol.
This last aspect is what allows us to not have to explicitly model any of the protocol-specific logic, as it was
done in prior work~\cite{tcp-paper}.
To quantify this aspect, our instrumentation of the reference implementation of the TCP protocol only required an additional 300 lines of mostly boilerplate code (which we envision can be automated in the future), while the 
mapper implemented~\citet{tcp-paper} to model the concretization function consisted of
2,700 lines of code modeling the complex logic of TCP.
For QUIC, our instrumentation required just over 2,000 lines of again mostly boilerplate code. Given the complexity of QUIC (which involve cryptography and other complex features), it is hard to imagine replicating the approach 
from~\cite{tcp-paper}.

To make the approach more effective, we introduce a few optimizations.
First, whenever the reference implementation sends new packets (as part of an intermediate step) that do not match
our abstract query, we store such packets in a list and use them to try to answer future queries.
That is, when an abstract query comes in, we first look into this list to check if any of the packets has the desired
abstract value and, if this is the case, we send that packet to the \implementation.
Second, we cache all intermediate pairs between abstract I/Os and concrete I/Os in a data structure called an \ot $O \subseteq \{ (a, c) \mid a \in (\widehat\Sigma^* \times \widehat\Gamma^*), c \in (\Sigma^* \times \Gamma^*) \}.$

In Section~\ref{sec:learning}, we use this cache to synthesize
richer models that go beyond finite abstract alphabets and can capture concrete packet numbers
and other numerical quantities.
We keep track of received packets for each query and also use this record to detect retransmitted packets that should not be part of the response due to nondeterminism.

Now that we have constructed an abstraction that produces a simplified abstract  alphabet, and we have effective ways of querying over this alphabet, we have built the interface necessary to interact with the \sul module. We could then, for example perform our 3-way TCP handshake by sending the input trace $\texttt{SYN(?,?,0)}$ \linebreak $\texttt{ACK(?,?,0)}$ and we would get the output trace $\texttt{ACK+SYN(?,?,0)}$ $\texttt{NIL}$, which accurately represents the flags of our 3-way TCP handshake. However, because it uses our abstract alphabet the \adapter cannot tell us the exact sequence or ack numbers it picked as this would cause nondeterminism.

%% file: tex/process.tex
We now describe how the learning module interacts with the \sul to learn models of the implementation.
In Section~\ref{sec:mat-interface}, we formally define the queries the learner is allowed to ask to construct a model.
In Section~\ref{sec:mealy}, we recall existing synthesis techniques for learning Mealy Machines (i.e., automata with outputs).
In Section~\ref{sec:Synthesis} we use program synthesis to learn a more detailed model that is capable of recovering register values and changes, like sequence and acknowledgement numbers in the TCP protocol.

\subsection{Learner Interface}
\label{sec:mat-interface}

Thanks to the techniques presented in Section~\ref{sec:sul},
the \sul can be treated as a \emph{query oracle}  that can answer the question "\textit{If I send this input sequence, what will the implementation return?}". With this deterministic query oracle we have all we need to create an interface for many model-learning algorithms for finite state machines.
In particular,   we can use the query oracle to implement (or at least approximate) two types of queries that a learner can ask:
\begin{description}[leftmargin=*]
    \item[Membership Queries:] $a_O = mq(a_I)$? where $a_I \in \widehat\Sigma^*$ and $a_O \in \widehat\Gamma^*$. These are single traces $a_I / a_O\in\widehat\Sigma^*\times\widehat\Gamma^*$ of the system that give the \learner  knowledge about the specific traces produced by the \sul so that it can build a hypothesis model $H$.
    \item[Equivalence Queries:] $eq(H)$? where $H$ is a hypothesis model the \learner believes to be correct and wants to know if it is equivalent to the \sul; the answer is either $a_I / a_O$, a counterexample trace that distinguishes the \sul from the hypothesis $H$, or no counterexample is returned, and the model is considered a correct abstraction of the \sul, terminating the learning process.\end{description}

In practice, Equivalence Queries require an oracle omniscient of the \sul, and if we had that, we would not have to learn the system in the first place. Instead, we can use heuristic Equivalence Oracles such that when a counterexample is returned, it is guaranteed to be a valid counterexample, but the absence of a counterexample no longer guarantees equivalence. This approach still gives us approximation guarantees---i.e., the  model is accurate with high probability with respect to the set of inputs
we use to test equivalence. It is now a good point to remind the reader that the goal of this paper is to unveil and discover potential
incorrect behaviors of the \sul, rather than provide behavior guarantees. Even if the learned models might not be 100\% accurate, they will still be helpful to 
analyze and detect anomalies as we will show in our evaluation.

Learners that depend only on these two types of queries were first studied in~\cite{angluin-learning-1987}
to learn deterministic automata and have since been extended to many types of state machines.

\subsection{Learning Mealy Machines}
\label{sec:mealy}

With oracles capable of answering membership and equivalence queries, we can use existing algorithms~\cite{learnlib} to learn 
 Mealy machines of the abstract behavior of the \sul. 
 Intuitively, a Mealy machine is a finite automaton that for every input symbol it reads, it also produces an output.

\begin{definition}[Mealy Machine]
A Mealy Machine is a tuple \linebreak $(S, S_0, \widehat\Sigma, \widehat\Gamma, T, G)$, such that:
 $S$ is a finite set of states, $S_0 \in S$ is the initial state, $\widehat\Sigma$ is the abstract input alphabet, $\widehat\Gamma$ is the abstract output alphabet, $T$ is the transition function $T \colon S \times \widehat\Sigma \to S$, and $G$ is the output function $G \colon S \times \widehat\Sigma \to \widehat\Gamma$. 
\end{definition}

\begin{example}
The Mealy machine in Fig.~\ref{TCP-Handshake}(b) is a model of the TCP 3-way handshake over input and output alphabets:

{\small\vspace{-.2cm}
\begin{align*}
	\widehat\Sigma & = \{\texttt{SYN(?,?,0)},\ \texttt{ACK(?,?,0)}\}	\quad
	\widehat\Gamma  = \{ \texttt{ACK+SYN(?,?,0)},\ \texttt{NIL}\}
\end{align*}
}
Given the input sequence $[\texttt{SYN(?,?,0)}, \texttt{ACK(?,?,0)}]$, this machine outputs the sequence
$[\texttt{ACK+SYN(?,?,0)}, \texttt{NIL}]$---starts in state $s_0$, when reading $\texttt{SYN(?,?,0)}$ it transitions to state $s_1$ and outputs $\texttt{\texttt{ACK+SYN(?,?,0)}}$, and then reading
$\texttt{ACK(?,?,0)}$, it transitions to $s_2$ and outputs $\texttt{NIL}$ (no packet).

\end{example}

Mealy machines have been studied extensively and there are many algorithms that can learn them using membership and equivalence oracles~\cite{learnlib}. At the high level, these algorithms issue membership queries to discover the behavior of the machine until they
can find a machine that is consistent---i.e., it correctly matches all the traces for which it has issued membership queries.
At this point, the algorithm issues an equivalence query, which can either end the learning process (in case of a yes answer) or 
cause the learning process to ask more membership queries and repeat this process.

\newcommand{\ACK}{\texttt{ACK}}
\newcommand{\SYN}{\texttt{SYN}}
\newcommand{\NIL}{\texttt{NIL}}

\begin{figure*}[t!]
    \centering
     \begin{subfigure}[b]{0.49\textwidth}
    \begin{tikzpicture}[node distance = 4cm, auto, transform shape]
        
        \node[state] (s0) {$s_0$};
        \node [above of=s0, node distance = .8cm] (i) {} ;
        \node[state, right of=s0] (s1) {$s_1$};
        \draw   (i) edge (s0)
          (s0) edge[loop left] node{\begin{tabular}{c}
			\tiny \texttt{ACK}(sn, an, 0) / \texttt{NIL} \\[-1ex]
			\tiny $r = \textbf{u}_1, pr = \textbf{u}_2, pi = \textbf{u}_3$ \end{tabular}} (s0)
            (s0) edge[bend left=5, above] node{\begin{tabular}{c}
			\tiny \texttt{SYN}(sn, an, 0) / \texttt{ACK}$(\textbf{o}_1, \textbf{o}_2,0)$  \\[-1ex]
			\tiny $r = \textbf{u}_4, pr = \textbf{u}_5, pi = \textbf{u}_6$ \end{tabular}} (s1)
            (s1) edge[bend left=5, below] node{\begin{tabular}{c}
			\tiny \texttt{SYN}(sn, an, 0)  / \texttt{NIL} \\[-1ex]
			\tiny $r = \textbf{u}_7, pr = \textbf{u}_8, pi = \textbf{u}_9$ \end{tabular}} (s0)
;
  \end{tikzpicture}
         \end{subfigure}
 	\centering 
 	\unskip\ \vrule\
     \begin{subfigure}[b]{0.49\textwidth}
    \begin{tikzpicture}[node distance = 4cm, auto, transform shape]
        \node[state] (s0) {$s_0$};
                \node [above of=s0, node distance = .8cm] (i) {} ;
        \node[state, right of=s0] (s1) {$s_1$};
              \draw   (i) edge (s0)
                 (s0) edge[loop left] node{\begin{tabular}{c}
			\tiny \texttt{ACK}(sn, an, 0) / \NIL  \\[-1ex]
			\tiny $r = r + 1, pr = pr, pi = \text{sn}$ \end{tabular}} (s0)
            (s0) edge[bend left=5, above] node{\begin{tabular}{c}
			\tiny \texttt{SYN}(sn, an, 0) / \texttt{ACK}$( pr, pr+1,0)$  \\[-1ex]
			\tiny $r = pr, pr = pr, pi = pi$ \end{tabular}} (s1)
            (s1) edge[bend left=5, below] node{\begin{tabular}{c}
			\tiny \texttt{SYN}(sn, an, 0)  / \NIL \\[-1ex]
			\tiny $r = r+1, pr = pr, pi = pi$ \end{tabular}} (s0)
;
  \end{tikzpicture}
         \end{subfigure}
              \caption{Extended machine with unknown terms (left) and corresponding synthesized machine  (right). \label{fig:learned-tcp}}     
\end{figure*}
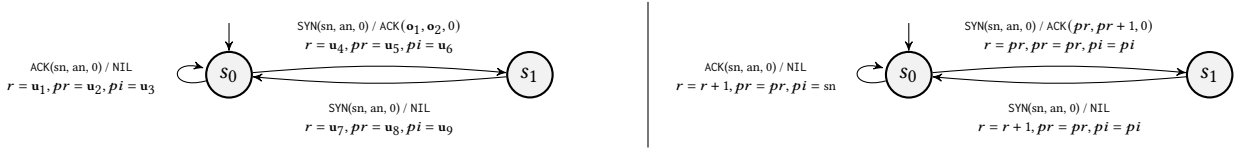

\prognosis uses the TTT algorithm~\cite{learnlib} which is guaranteed to learn a Mealy machine in time
polynomial in the size of the machine.
For example, when we run TTT on the TCP implementation using the abstract alphabet in~\Cref{ex:abstractalph}, TTT learns the
 model in Fig.~\ref{TCP-Handshake}(b) (we depict only transitions relevant to the handshake, though the learned model is deterministic and total).

\subsection{Synthesizing Rich Models}
\label{sec:Synthesis}

Mealy Machines can only model operations involving finite alphabets and cannot reason about numerical values---e.g., sequence numbers.
To capture the packet exchange in \Cref{TCP-Handshake}(a) we need not only the TCP flags currently learned in the model  depicted in \Cref{TCP-Handshake}(b) but also certain quantities\footnote{
Existing extensions of the learning algorithm that handle automata with counters~\cite{learnlib} do not meet the needs described in this paper as they do not support complex comparison operations and updates.}---sequence and acknowledgement numbers. 

In this section, we present an extension of Mealy machines that  adds registers, and numerical inputs and outputs.
We then show how the ideas in Section~\ref{sec:mealy} can be combined
with constraint-based synthesis techniques to learn these extended models.
\newcommand\upd[1]{\mathbf u(#1)}
\newcommand\out[1]{\mathbf o(#1)}
 While we could consider a wide range of enhancements to traditional automata,
we limit ourselves to extending automata to read and write integer values from packets,
and to increment or set to input values one of a finite number of registers $\vec x$.
These extensions capture common features of the protocols we are interested in.
A transition of an extended Mealy machine looks like the following:
$$ p \xrightarrow[]{\substack{I(\vec{i})/O(\out{\vec{x}}) \\ \vec{x} = \upd{\vec{i},\vec{x}}}} q$$
Informally, if the machine is in state $p$ and reads an abstract symbol $I$, possibly parametric on concrete
numerical values $\vec{i}$, it updates the registers $\vec x$  
 with values determined by $\upd{\vec i,\vec x}$, 
and it outputs an abstract symbol $O$ parametric on the values determined by $\out{\vec{x}}$. 
The update function $\upd -$ can take on quite complex values, but in the following 
we consider each register is updated with either a copy of a register, or an input value, or one of these incremented by 1.   
Similarly, the output function $\out -$ can, for each parameter, output the value of a register or that
value plus 1.

We start from a Mealy machine where the register updates and 
outputs are missing (as in \Cref{fig:learned-tcp}(left)) and our goal is to find concrete terms $\textbf{u}_1, \cdots, \textbf{u}_9$, and $\textbf{o}_1, \textbf{o}_2$ for each transition that result in an extended machine correctly modelling the \sul with respect to a given set of concrete traces.

We use the set of traces cached while learning the Mealy machine in the \ot $T$. For each pair $(a,c)\in T$ of abstract and concrete traces, 
we identify what path of the extended Mealy machine is traversed by the abstract trace $a$
and use the concrete trace $c$ to generate the constraints needed to identify the missing terms 
of each transition in the path---i.e., the terms that make the extended machine consistent with the concrete trace $c$.
If needed, the algorithm can solicit more example traces and add them to $T$. 
The constraints are then solved using an SMT solver and the solution is used to generate the needed terms.

Let us illustrate how we would synthesize the extended machine in \Cref{fig:learned-tcp}(right) from the the sketch in \Cref{fig:learned-tcp}(left). We consider the following concrete trace appearing in $T$:
\linebreak
$[(\ACK(0, 3,0)/\NIL), (\SYN(2, 5,0)/ \ACK(4, 5, 0))]$.
This time, we want to model that each symbol in the concrete trace also carries the synchronization number (sn) and the acknowledgement number (an)---e.g., for the first input these values are 0 and 3, respectively.
Given this trace, our algorithm generates constraints containing a number of variables used to denote what we are trying to
synthesize and what it means for the solution to be correct with respect to the input trace.
For each unknown term,
we have a finite list of possible terms we can instantiate it with.
For example, the unknown $\textbf{u}_1$ can be instantiated with one of the
8 terms in the list:
$[r,r+1,pr, pr+1,pi,pi+1,\textnormal{sn},\textnormal{an}]$.
In our constraints, we use an integer variable $E_{\textbf{u}_{1}}$ to indicate the possible choices (indices start at 0).
For example, $E_{\textbf{u}_{1}}=1$ indicates that  the
term $r+1$ will be the solution for the unknown $\textbf{u}_1$.

As register values will change for each trace, 
our constraints need to model how the values of the register are
updated throughout the execution.
To do so, we introduce variables that track the values of each
register after reading the $i$-th input in the trace.
For example, $r[i]$ indicates the value stored in register $r$
after reading the first input packet. 
When generating constraints for multiple traces, we will have a variable
$r_\pi[i]$ for each trace $\pi$ and index $i$.

The following simplified set of constraints capture
the synthesis problem we are trying to solve:

{\footnotesize\vspace{-.25cm}
\[
\begin{array}{ll}
 \multicolumn{2}{l}{\color{ForestGreen}\text{// Constraints for $\textbf{u}_1$, $\textbf{u}_2$, $\textbf{u}_3$, $\textbf{u}_4$, $\textbf{u}_5$, $\textbf{u}_6$}} \\
 \multicolumn{2}{l}{0 \le E_{\textbf{u}_1} \le 7 \text{\color{ForestGreen}\quad // The value of $E_{\textbf{u}_1}$ encodes the 8 possible terms for $\textbf{u}_1$}}  \\
 E_{\textbf{u}_1} = 0 \implies r[1] = r[0] 
& E_{\textbf{u}_1} = 1 \implies r[1] = r[0]+1 \\ 
 E_{\textbf{u}_1} = 2 \implies r[1] = pr[0] 
& E_{\textbf{u}_1} = 3 \implies r[1] = pr[0]+1 \\
 E_{\textbf{u}_1} = 4 \implies r[1] = pi[0] 
& E_{\textbf{u}_1} = 5 \implies r[1] = pi[0]+1 \\
 E_{\textbf{u}_1} = 6 \implies r[1] = 0  
& E_{\textbf{u}_1} = 7 \implies r[1] = 3 
\\ \ldots \\
 \multicolumn{2}{l}{\color{ForestGreen}\text{// Constraints for $\textbf{o}_1$, $\textbf{o}_2$}} \\
 \multicolumn{2}{l}{0 \le E_{\textbf{o}_2} \le 5 
 \text{\color{ForestGreen}\quad // The value of $E_{\textbf{o}_2}$ encodes the 6 possible terms for $\textbf{o}_2$}}  \\
 E_{\textbf{o}_2} = 0 \implies r[2] =5  
& E_{\textbf{o}_2} = 1 \implies r[2] + 1=5 \\
 E_{\textbf{o}_2} = 2 \implies  pr[2] =5
& E_{\textbf{o}_2} = 3 \implies  pr[2] + 1=5 \\
 E_{\textbf{o}_2} = 4 \implies  pi[2]  =5
& E_{\textbf{o}_2} = 5 \implies pi[2] + 1=5 \\ 
\end{array}
\]\vspace{-.25cm}}

Note that setting $E_{\textbf{u}_1}=7$ corresponds to selecting the term an (i.e., the input {\ACK} number)
as a solution to the unknown $\textbf{u}_1$.
In this case, the constraints  model that the value of the register $r$ after reading the \textit{first} packet (i.e., $r[1]$)
should be equal to the {\ACK} number of the \textit{first} input packet (i.e. 3). 
Similarly, setting $E_{\textbf{o}_2}$ to the value $4$ corresponds to selecting
the term $pi$ as a solution to the unknown $\textbf{o}_2$.
In this case, the constraints model that the value of the register $r$ after reading the \textit{second} packet (i.e., $r[2]$)
should be equal to the {\ACK} number of the \textit{second} output packet (i.e., 5).

Repeating this process on more examples, e.g  \linebreak $[(\SYN(2, 3, 0)/ \ACK(4, 5, 0)), (\SYN(2, 3, 0)/ \NIL)]$, yields the automaton in \Cref{fig:learned-tcp}(right).
This automaton corresponds to the solution that selects 
$E_{\textbf{u}_1}=1$ and $E_{\textbf{o}_2}=3$.

While this section shows a particular type of model, our framework is general and allows to implement other synthesis and learning algorithms for more complex models (e.g., allowing more complex constraints and updates) due to its interaction with the \sul.

Sometimes the synthesized models can contain incorrect register patterns if the pattern is not completely covered by the \ot traces. These are detected through random equivalence testing, and trigger new queries in the synthesis algorithm---i.e.,
the synthesizer will restart with a larger set $T$ of traces to learn for as well as negative example---i.e., traces
that the model should not contain. The constraints discussed in this section can be easily adapted to handle
negative examples.

%% file: tex/checking.tex
\section{Analysis Module}
\label{sec:Model-Analysis}

The last module of \prognosis enables analysis techniques using the outcomes of the learning modules to help
the user infer behaviors of the \sul.
Different abstraction levels allow us to expose different types of anomalies, however the analysis module is limited to uncovering logic errors captured in the model, which is restricted to observable events captured in the learning process. Some more nuanced quantity specific bugs can be analysed through the synthesized model, however these are limited to linear patterns.

We focus on analysing models for bug finding and knowledge acquisition rather than providing verified guarantees due to not all abstraction levels capturing enough information to fully verify the specification. In this section, we present some of these analyses and how they can be used to identify undesired behaviors.

\subsubsection*{Nondeterminism Check}
As explained in \Cref{remark:nd}, the learning module typically expects deterministic behavior (i.e., the same input trace should trigger the same output trace). 
In general, nondeterminism can arise in our system for two reasons: 
(1) the abstract alphabet is so simplified that concrete traces corresponding to different behaviors are collapsed into the same input trace, or
(2) the implementation is producing nondeterministic outputs in cases where it should not.
For every  \learner query, \prognosis expects a deterministic answer, as it is attempting to build a deterministic model. However, due to the nature of active-learning, environmental events such as latency and packet loss could cause non-determinism to be observed. To avoid this \prognosis will execute the query a specified minimum number of times, and if these are not all the same, it will continue to do so until a specified percentage of certainty is achieved. If after a limit number of queries nondeterministic is discovered, the learning process pauses, and the \adapter verifies the cause of the nondeterminism. If reason (1) is the cause, the user will see that the abstraction is too coarse and can provide a richer abstraction. While it is possible to use more complex learning algorithms that can handle nondeterminism in the traces, we argue that, because of  reason (2), detecting sources of nondeterminism is a powerful analysis technique that can shed light on some undesired behaviors.
As we show in Section \ref{sec:nondeterminism-bug}, \prognosis was able to unveil a complex bug in a QUIC implementation thanks to
the nondeterminism check.

\subsubsection*{Learned Model Analysis}

If learning succeeds, \prognosis can produce visualizations of the models which can help a user understand whether an implementation works as expected. In addition, \prognosis also exposes various algorithms and decision procedures to check 
properties of the learned models.
For example, \prognosis can check equivalence of models of two different implementations of the same protocol
 (whether they accept the same input/output traces). For Mealy machines, there are algorithms for
performing this equivalence check efficiently~\cite{10.1145/322017.322020}.
 \prognosis can produce concrete example traces that show the difference between the behaviors of the two 
implementations.
 We show in Section \ref{sec:Quic-Tracker-Bug} how \prognosis is able to unveil a complex bug in a QUIC implementation thanks to these equivalence checks, and the visualizations of the differences were instrumental in communicating the problem to the developers.

\prognosis also allows the user to specify temporal properties in logics such as LTL or CTL, e.g., {\em Packet numbers are always increasing}, and check whether the models satisfy such properties. For Mealy machines, this procedure simply boils down to checking that the traces of the model are a subset of those
allowed by the property, a decidable problem~\cite{10.1145/322017.322020}.
For extended machines with counters, this problem is in general undecidable and we rely on randomised testing to make the analysis practical.

%% file: tex/evaluation.tex
We evaluate \prognosis  using two case studies.
First, we show that \prognosis can replicate prior works on learning 
models of TCP implementations~\cite{tcp-paper} (\Cref{Sec:case1}). 
Second, we use \prognosis to learn models of different implementations of the IETF QUIC protocol, which is undergoing standardization after successful initial development at Google (\Cref{Sec:case2}).
Overall, we evaluate \prognosis' effectiveness in identifying unintended behavior in real protocol implementations, as well as its modularity and reusability.

 \prognosis was implemented using several programming languages. The abstract learner is implemented in 3,500 lines of Java and uses the automata learning library LearnLib~\cite{learnlib}. The extended Mealy machine synthesizer is implemented in 300 lines of Python and uses the Z3 SMT Solver\cite{de-moura-z3-2008}.
 
 \subsection{Learning a TCP Implementation}
\label{Sec:case1}

In this case study, we demonstrate how \prognosis can reproduce results from previous work on learning models 
of TCP implementations~\cite{tcp-paper}.  We learn the \emph{Ubuntu 20.04.1 LTS} TCP stack, particularly kernel version \texttt{5.8.0-40-generic}. 

We provide \prognosis with the same abstract alphabet $\widehat\Sigma$
and translation pair $(\alpha, \gamma)$ used in prior work~\cite{tcp-paper}. 
Concretely, the abstract alphabet models the TCP flags for the packet, 
leave the sequence and acknowledgement numbers unspecified, and provide the payload length.

{\footnotesize\vspace{-.3cm}
\begin{align*}
\widehat\Sigma = \{ & \texttt{SYN(?,?,0)},\texttt{SYN+ACK(?,?,0)}, 
                 \texttt{ACK(?,?,0)}, 
                 \texttt{ACK+PSH(?,?,1)},\\ 
               &  \texttt{FIN+ACK(?,?,0)},
                  \texttt{RST(?,?,0)}, 
                \texttt{ACK+RST(?,?,0)} \}
\end{align*}\vspace{-.4cm}}

To implement the abstraction function $\alpha$ and the concretization function $\gamma$ we need a reference implementation.
To illustrate the flexibility of \prognosis, we use the mapper implementing the TCP logic designed in~\cite{tcp-paper}
as reference and instrument it. Note that we could have used any other TCP implementation
for this task. Modifying the reference implementation required only 300 lines of code to integrate with
\emph{Scapy}~\cite{noauthor-scapy-nodate} instead of the >3000 lines of code
required to implement a concretization function in~\cite{tcp-paper}. 

The model learned by \prognosis has 6 states and 42 transitions and took 4,726 membership queries to learn. This model is slightly different to the one learned by~\citet{tcp-paper} due to major differences made to the implementations over time. 
Unfortunately, despite multiple attempts, we were not able to run the prior technique by~\citet{tcp-paper} on the new TCP implementation and we cannot see what model it will learn now.
This case study and the difficulty of adapting prior work to new implementations showed that we could reproduce the level of detail
obtained in prior work with very minor effort thanks to the modular components of \prognosis.

While we do not perform any analysis in this case study, we show that our synthesis procedure can help us
derive richer models than those considered in prior work. In particular, using the technique described in \Cref{sec:Synthesis}, we could synthesize
a model describing the behavior of extra parameters such as the \texttt{dataOffset}.
Remarkably, applying this richer analysis only required changing one line of code in the implementation---simply selecting   which concrete values of the TCP implementation we wish to synthesize over.

\subsection{Learning QUIC Implementations}
\label{Sec:case2}

In this section, we analyze 4 implementations of the QUIC protocol. We devise the abstract alphabet $\widehat\Sigma$, provide a reference implementation for the translation pair $(\alpha, \gamma)$, and a set of properties from the QUIC specification to analyze. We show how \prognosis detected behavior that led to a change in the QUIC specification, as well as identified bugs in real implementations of the QUIC protocol.

\subsubsection{Background}
The QUIC protocol~\cite{thomson-quic-nodate} combines specific features of independent protocols that are commonly used together, reducing the overhead that arises from having to coordinate these features. Specifically, QUIC is optimized for the Web, and as such, collapses the services provided by HTTP, TLS, and TCP into a single \textit{super protocol}. 
Therefore, QUIC is extremely complex, and its design has been carefully thought out to ensure these services that were previously isolated can be communicated efficiently. QUIC achieves this goal using encapsulation.
In TCP, a single packet is enough to communicate everything needed, from specific signals like \texttt{SYN} or \texttt{ACK} to the payload of the application layer itself. In QUIC, both signaling, and data transmission happen in \emph{frames}. A packet serves merely as a mean of safely transporting different types of frames.  

In total, QUIC provides 7 packet types and 20 frame types, each responsible for signaling a specific aspect of the protocol. We will not be diving into the details here, due to their complex nature, and instead, we will introduce specific frames and packets as needed to explain key properties of the protocol and defer the reader to the QUIC specification~\cite{thomson-quic-nodate} for further details. We will consider the following QUIC implementations:
\begin{description}[leftmargin=*]
    \item[Quiche]is Cloudflare's QUIC implementation~\cite{cloudflare-quiche} that allows QUIC connections to any website protected by Cloudflare network. 
    Cloudflare's CDN supports $>25$ million websites.
    \item[Google's QUIC] implementation~\cite{google-quic} runs on Google servers and on Chromium browsers. Since October 7 of 2020, 25\% of Google Chrome users had QUIC support enabled by default, with that proportion increasing over the followed weeks. 
    \item[Facebook's mvfst] QUIC implementation~\cite{facebook-mvfst} is used mainly in the Proxygen server implementation. This server implementation is responsible for powering most public facing connections to facebook.com, as well as API connections used by the Facebook and Instagram mobile apps.
    \item[QUIC-Tracker] is an implementation~\cite{piraux-observing-2018} designed to run testing scenarios over other implementations, with the goal of testing what technologies different implementations support.
\end{description}
Because these first three implementations account for a large portion of the web traffic, 
identifying erroneous behaviors in any of them is of critical importance.

\subsubsection{Learning Models of QUIC Implementations}
\label{sec:models-learned}
We use \prognosis to learn models of the first three  QUIC implementations described above and analyze a subset of the properties from IETF's Draft 29 \cite{thomson-quic-nodate}---e.g., {\em The sequence number on each newly-issued connection id must increase by 1} and {\em An endpoint must not send data on a stream at or beyond the final size}.
\prognosis requires 3 things from the user: An abstract input alphabet, a translation pair $(\alpha, \gamma)$ to be able to convert between abstract and concrete alphabets, and optionally, a set of properties $\Phi$ that we would like to test. We use an abstract input alphabet $\widehat\Sigma$ with 7 symbols:

{\footnotesize \vspace{-.3cm}
\begin{align*}
\widehat\Sigma = \{ & \texttt{INITIAL(?,?)[CRYPTO]}, 
	             \texttt{INITIAL(?,?)[ACK,HANDSHAKE_DONE]}, \\
	& \texttt{HANDSHAKE(?,?)[ACK,CRYPTO]}, \\
	& \texttt{HANDSHAKE(?,?)[ACK,HANDSHAKE_DONE]}, \\
	& \texttt{SHORT(?,?)[ACK,MAX_DATA,MAX_STREAM_DATA]}, \\
	& \texttt{SHORT(?,?)[ACK,STREAM]}, 
	 \texttt{SHORT(?,?)[ACK,HANDSHAKE_DONE]} \}
\end{align*}\vspace{-.5cm}}

The first four symbols are used to open the connection, perform the handshake, and transmit data, and the last three
symbols allow us to model properties related to flow control. We focus on this alphabet, as due to QUIC's numerous packet and frame types, choosing to learn the behavior of, for example, all packet types containing up to 3 frames,  would already give us an alphabet with over 30,000 symbols. Although, in theory, we could learn over such large alphabets, the learning process would take an infeasible amount of time. As we will show in the rest of the section, these seven symbols 
are enough to capture the main connection establishment, handshake, data transmission, and flow control behavior.

To build the translation pair $(\alpha, \gamma)$ in the \adapter, we use \emph{QUIC-Tracker}~\cite{piraux-observing-2018} as a reference implementation and instrument it to convert between abstract and concrete symbols. Specifically, we have implemented the format we picked for our abstract alphabet, and whenever our \adapter receives an abstract request from the learner, we use QUIC-Tracker's state to determine if there is a valid concrete packet matching the abstract packet that should be sent from the current state. If so, we send this queued packet, if not, we request that QUIC-Tracker sends a new packet matching our abstract requests, filling in the concrete values required to make it a valid packet under the current state. QUIC-tracker then encodes and sends 
the created packet to the \implementation we are learning a model for and waits for a response. Once this response is detected, it is abstracted and sent back to the learner as a response.

 The QUIC \adapter is comprised in total of over 10,000 lines of Go for its reference implementation, however the instrumentation requires only an additional  2,000 lines of code, 
which we believe is a relatively low cost for \prognosis to handle a protocol as complex as QUIC.
Furthermore, the same instrumentation can be used to learn models of all the QUIC implementations.
When ran on these three implementations \prognosis could learn models for two of the three implementations (see \Cref{sec:nondeterminism-bug} for explanation). The two models had 12 and 8 states, and 84 and 56 transitions, respectively. Learning took 24,301 queries for one implementation and 12,301 queries for the other. Having these two models enables reasoning with much fewer traces: for the alphabet $\widehat\Sigma$ above there are 329,554,456 traces of length up to 10, however we only need to check 1210 and 715 of those traces, respectively, for the two implementations. This reduction is gained because the traces in the learned model are a subset of all traces that can be made with the alphabet.

In the rest of the section, we describe what issues were unveiled in the QUIC RFC and implementations using a number of formal and informal analysis of the models generated by \prognosis. 
As some of the problems we have identified introduce serious security vulnerabilities that the developers of
the protocols have not yet addressed, we will not identify in which specific implementations the issues were found.

\subsubsection{Issue 1: RFC Imprecision}
\label{sec:RFC-Improvements}

\paragraph{\bf Analysis technique and outcome:}
When comparing the models learned for different implementations, we discovered that these models, as reported in Section~\ref{sec:models-learned}, had vastly different sizes.
We manually explored the models to identify the key differences in structure and 
found that different implementations were not consistent on what to do if a client resets the \emph{Packet Number Spaces} 
when retrying a connection. We reported this anomaly to the IETF QUIC Working Group who issued a fix~\cite{noauthor-let-nodate} on the next version of the specification to make the expected behavior clearer.

\paragraph{\bf Underlying issue:}
As QUIC is a protocol undergoing standardization, its RFC is a living document that is still being perfected. One of the guiding principles for this document is the Robustness Principle~\cite{postel-dod-1980} that was already employed in the development of TCP. The principle states that a system should be strict about what it sends, and liberal on what it receives. As such, the QUIC RFC is perhaps a looser specification than those used for formal verification, 
and certain aspects of the specification are ambiguous. 

Our reported inconsistency unveiled an ambiguity in the behavior of \texttt{RETRY} packets. \texttt{RETRY} packets are a special kind of packet used by a server to verify the source address of the received packet. It is called \texttt{RETRY} as the client should retry to open the connection with the \texttt{RETRY} token received from the server. 
Most importantly, our \textbf{report led to a discussion} on the topic by the RFC maintainers, and \textbf{a fix was issued} to clarify that a server \emph{MAY} abort the connection when a client resets their Packet Number Spaces \cite{noauthor-let-nodate}.

\subsubsection{Issue 2: Nondeterminism in Connection Closure}
\label{sec:nondeterminism-bug}

\paragraph{\bf Analysis technique and outcome:}
As described in \Cref{sec:Model-Analysis}, \prognosis ensures that the answer to every  \learner query  has a deterministic response via the nondeterminism check. During this check, we found that it was possible to have Facebook's mvfst
 close the connection and remain in a state where it will not always respond with \texttt{RESET}s to subsequent packets.
\paragraph{\bf Underlying Issue}
For backwards compatibility reasons, QUIC packets are carried over the UDP protocol. While TCP has been vastly optimized in devices ranging from internet middle-boxes such as switches and routers to endpoint devices such as servers and clients, UDP has not received the same level of attention, mostly due to TCP being the \textit{de facto} protocol for the transport layer. 

To understand this bug, we need to introduce a specific QUIC frame: \texttt{HANDSHAKE_DONE} is a signaling frame sent by the server (and only the server) to notify the client that the handshake is now complete, and it can proceed to transmit its data. We include this frame in the input alphabet used by the \learner, and as such the learned model will depict what  happens in case the client sends this server-only frame at different states of the system. The ideal response is that, as this frame should never be sent by a client, the server would treat receiving this frame as a protocol violation error, and immediately close the connection, and any further packets sent on the same connection \textit{may} be met with a \texttt{RESET} packet, a decision that is up to the developer to make. The \texttt{RESET} packet type is a last case resort that an endpoint can use to notify the other side that this connection no longer exists, even when data transmissions keys are no longer available.

\prognosis detected that if the client starts a connection as usual with \texttt{INITIAL(?,?)[CRYPTO]}, and instead of finishing the handshake, sends a \texttt{HANDSHAKE_DONE} frame to the server with the \texttt{HANDSHAKE(?,?)[ACK,HANDSHAKE\_DONE]} abstract symbol, this packet will be met with a \texttt{CONNECTION_CLOSE} frame, effectively closing the connection. However, after testing the same packet sequence repeatedly, it found that if the client keeps sending packets, it is only in 82\% of the responses that following packets are met with a \texttt{RESET}.

This behavior discovered in our analysis is \textbf{erroneous}: a specific implementation may choose to use \texttt{RESET} packets, but it must be \textbf{consistent} in the decision. It cannot nondeterministically switch between sending and not sending packets. Furthermore, this RESET behaviour has no back-off mechanism, meaning that a client can exploit this bug to request new packets from the server on demand. 
The client could keep sending the exact same packet, with no computation needed, and the server would have to produce new \texttt{RESET} packets every time it receives this unexpected packet. This behavior, coupled with the fact that \texttt{RESET} packets are relatively small, means that system-level UDP optimizations are not triggered, resulting in sending each \texttt{RESET} packet being an expensive operation. If exploited, this behavior can be used in a \emph{Denial-of-Service} attack against the server. This critical bug has been acknowledged by the developers and will be fixed soon.

\subsubsection{Issue 3: Inconsistent Port on RETRY in QUIC-Tracker}
\label{sec:Quic-Tracker-Bug}

\paragraph{\bf Analysis technique and outcome:}
After learning the model of one of the implementations we detected a discrepancy (in terms of traces)
in how two different implementations handle the \textit{Retry Mechanism}: if the server sent a \texttt{RETRY} packet the model would transition to a state where connection establishment was impossible. This greatly contrasted with behavior learned in another model. 

One of the properties of a confirmed erroneous trace is that either the \implementation or the \adapter are behaving unexpectedly. In this case, it was the reference implementation, QUIC-Tracker, that had a bug in its retry mechanism, and only one of the target implementations considered it critical enough to not allow connection establishment to take place.

\paragraph{\bf Underlying Issue:}
QUIC's Retry Mechanism, as introduced in \Cref{sec:RFC-Improvements}, is a way for the server to validate that a packet is indeed being sent from a specific IP address and port, and not just an attacker replaying packets from a spoofed source. The client starts by sending a \texttt{ClientHello} as usual, but this time the server replies to it with a \texttt{RETRY} packet. This packet has a unique token which the client must take and send in a new \texttt{ClientHello} from the same IP address and port. The server then checks if the \texttt{Retry Token} matches the one previously sent to confirm these packets are being sent from the claimed source.
Our reference implementation was correctly returning the token; however it was doing so with a new UDP socket using a random free port instead of the one we were using before. As such, the token would be sent from a different port, and address validation would fail, interrupting the handshake. 

\subsubsection{Issue 4: Stream Data Blocked Bug in Google QUIC}
\label{sec:Google-Bug}

\paragraph{\bf Analysis technique and outcome:}
Using the extended models of \Cref{sec:Synthesis}, we synthesized an extended Mealy Machine (shown in Appendix \ref{sec:apx-synth}) that describes how the \texttt{Maximum Stream Data} field in the \texttt{STREAM_DATA_BLOCKED} frame changes over the states. By inspecting this model we were able to detect that this field always has the value \texttt{0}, and is never updated, even when the stream gets blocked. 
\paragraph{\bf Underlying Issue:}
One of the key features of QUIC is flow control. For this, QUIC makes use of a series of frames dedicated to coordinating flow limits between the two endpoints. One such frame, the \texttt{STREAM_DATA_BLOCKED} frame, is responsible for alerting the other endpoint when we would like to send data on a stream, but the flow control limits imposed on us by the endpoint itself do not allow us to do so. This frame has two fields, a \texttt{Stream ID} field indicating which stream is blocked, and a \texttt{Maximum Stream Data} field, indicating at what offset of the data the block happened. 
We found that although one of the implementations used these frames and did not transmit data over the agreed limits, it did not set the \texttt{Maximum Stream Data} correctly, and instead used the constant \texttt{0}. 
The developers have confirmed that this section of the specification was incorrectly implemented.
They explained that \texttt{0} was a placeholder set during initial development, which they had
forgotten to update.

%% file: tex/artifact.tex
\appendix
\section{Artifact Appendix}

\subsection*{Abstract}

We provide the source code and scripting to run \prognosis, as well as re-run and verify claimed properties and results. 

\subsection*{Scope}

The artifact covers the full code and tooling used in both the development and execution of \prognosis, as well as the source code of its targets. 

\subsection*{Contents}

The artifact bundles the learner, adapter, synthesizer, and target implementations required to run \prognosis, as well as custom scripting and documentation.

\subsection*{Hosting}
The artifact's code, scripts, and documentation are fully accessible at: https://doi.org/10.5281/zenodo.5040974

\subsection*{Requirements}

A device capable of running Docker and Docker Compose, as well as x86 based images, either natively or through virtualisation.

A minimum of 8 GB of memory, recommended 16 GB for running all experiments. 

%% file: tex/appendix.tex
\begin{appendix}

\section{Learned Mealy Machines}
\subsection{TCP Implementation}
\includepdf[pages={1}]{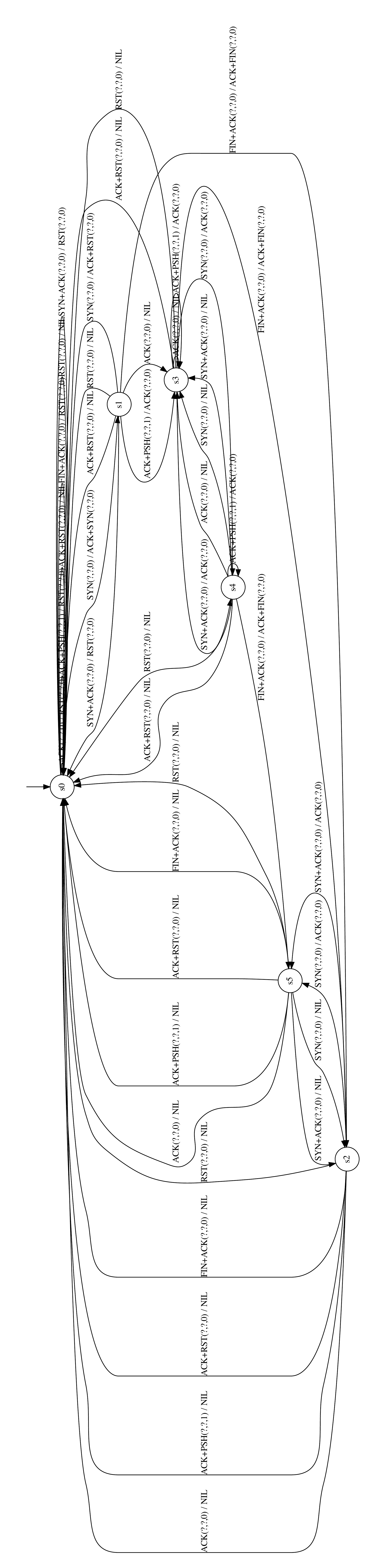}

\subsection{Google's QUIC}
\includepdf[pages={1}]{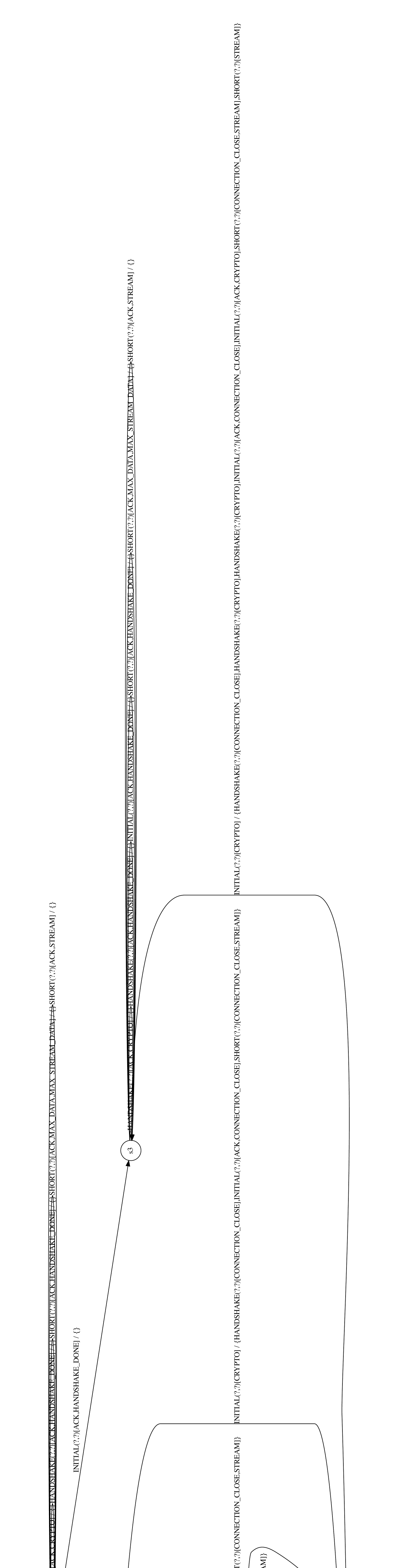}
\includepdf[pages={2}]{graphics/imp-2.pdf}
\includepdf[pages={3}]{graphics/imp-2.pdf}
\includepdf[pages={4}]{graphics/imp-2.pdf}

\subsection{Cloudflare's Quiche}
\includepdf[pages={1}]{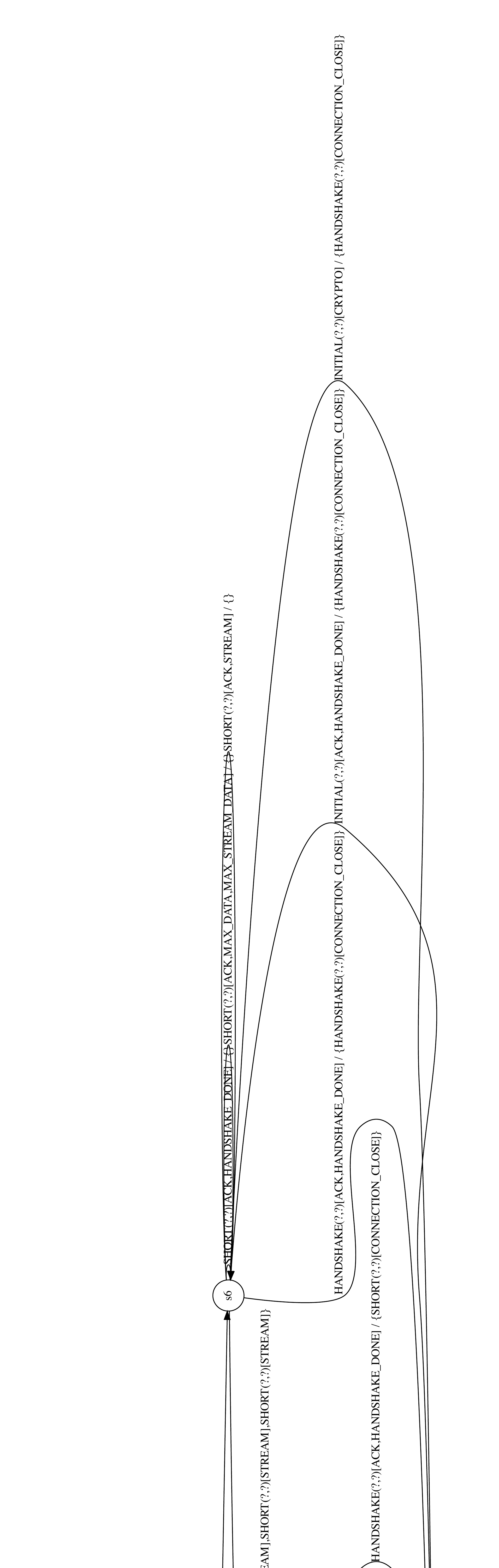}
\includepdf[pages={2}]{graphics/imp-3.pdf}
\includepdf[pages={3}]{graphics/imp-3.pdf}
\includepdf[pages={4}]{graphics/imp-3.pdf}

\section{Synthesized Extended Mealy Machines}
\subsection{Google's QUIC -- \texttt{Maximum Stream Data} Field} \label{sec:apx-synth}
\includepdf[pages={1}]{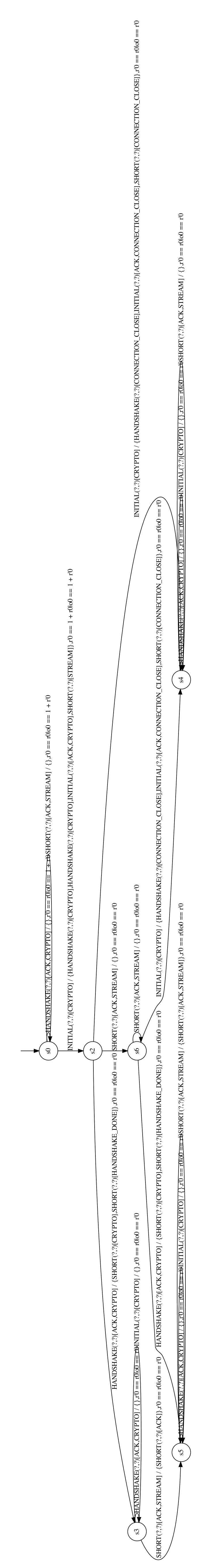}

\end{appendix}